\documentstyle[preprint,tighten,aps]{revtex}

\begin{document}
\title{Spectral patterns in the nonstrange baryon spectrum}
\author{P. Gonz\'alez$^{1}$, J. Vijande$^{1,2}$, A. Valcarce$^{2}$, H.
Garcilazo$^{3}$}
\address{$^1$ Departamento de F\' \i sica Te\'orica and IFIC,
Universidad de Valencia - CSIC,
E-46100 Burjassot, Valencia, Spain}
\address{$^2$ Grupo de F\'{\i}sica Nuclear and IUFFyM, Universidad de
Salamanca,
E-37008 Salamanca, Spain}
\address{$^3$ Escuela Superior de F\'\i sica y Matem\'aticas, 
Instituto Polit\'ecnico Nacional,
Edificio 9, 07738 M\'exico D.F., Mexico}
\maketitle

\begin{abstract}
We extract, from a quark model potential that reproduces the number
and ordering of nonstrange baryonic resonances up to 2.3 GeV, the
quantum numbers for the dominant configurations in the ground and first
non-radial excited states. From the pattern of quantum numbers we
identify, from data, spectral regularities that allow us to predict
the expected high spin low-lying spectrum from 2.3 to 3.0 GeV.
$N-\Delta$ degeneracies and $N$ parity doublets
showing up can be interpreted in terms of a simple dynamics.
\end{abstract}

\vspace*{2cm} \noindent Keywords: nonrelativistic quark models, baryon
spectrum, missing states \newline
\noindent Pacs: 12.39.Jh, 14.20.-c, 14.20.Gk

\newpage

\section{Introduction}

There has been much interest during the last years in the high energy part
of the hadronic spectrum. The aim is to get a better understanding of the
dynamics involved, in particular the confinement mechanism in hadrons. On
the theoretical side some progress has been made. On the one hand,
unquenched lattice QCD points out a string breaking in the static potential
between two quarks \cite{Bal01,Bal05} what should be properly incorporated
in the phenomenological description of the high energy hadronic spectrum
through the coupling to open decay channels. On
the other hand, the idea of a parity multiplet classification scheme at high
excitation energies as due to chiral symmetry has been suggested \cite{Jid00}
and recently put in question \cite{Jaf05}. On the experimental side the lack
of precise and complete data prevents, at the current moment, to establish
in a clear-cut way a classification scheme for the highly excited hadron
spectrum. Hence it may be appropriate to try to extract, from a simple
dynamical quark model calculation and its comparison to data, some spectral
patterns. To this purpose we shall consider the light baryon spectrum for
which an extensive collection of data including high spin state masses,
though not very precise in some cases, exists.

We shall use a nonrelativistic quark potential model. The application of
such approach for high excited states deserves some comments since the
quarks inside the baryons may be ultra-relativistic, $q\overline{q}$ pair
creation out of the vacuum becomes more and more relevant as the number of
open decay channels increases, etc. As it is well-known these shortcomings,
always implicit in the application of the naive quark model, do not
invalidate its usefulness to provide information about the pattern of
the quantum numbers of the baryon states. To get an unambiguous assignment
of quantum numbers one should require the model to generate, inasmuch as
possible, the correct number and ordering of the well established
experimental states. Different dynamical models providing accurate
descriptions of the low energy baryon spectrum have been proposed
\cite{Cap00}. Most of them rely on non-limited
(usually linearly dependent on the interquark distance) confinement
interactions. As a consequence, above 1.9 GeV, there are many more predicted
states than observed resonances. This {\it missing state}
problem can be solved by ascribing experimental
resonances (in the $\pi N$ partial wave analysis) to predicted
states with a significant coupling to the $\pi N$ formation 
channel \cite{Cap93}.

Alternatively the missing state problem can be obviated by using a
quark-quark screened potential as shown in Ref. \cite{Vij04}, where a correct
prediction of the number and ordering of the known $N$ and $\Delta$ 
resonances, up to 2.3 GeV mass, is obtained. This points out to
screening as an effective manner to give somewhat account of the coupling to 
$\pi N$ channels (lattice calculations have unveiled the close
relation between screening and the opening of decay channels). Such
interpretation is supported by the fact that the quantum numbers of the
dominant configurations (for the ground and first excited states of given
$J$) coming out from a screened potential (containing only confinement and a
residual one-gluon exchange (OGE) interaction) are in perfect
agreement with the ones ascribed to data from refined non-screened models
through the coupling analysis as we have explicitly checked for 
$J\leq 7/2$ (see Ref. \cite{Cap93} for the
identification of the available dominant non-screened configurations).

Though one cannot expect, from the minimal screened dynamics
employed, to get an accurate description of the values of the baryon masses,
it is clear its convenience to provide an unambiguous assignment of quantum
numbers to baryon states beyond the sometimes ad hoc arguments that have to
be used in non-screened models to identify
the experimental resonances when the $\pi N$ 
coupling strengths have similar values for different predicted
candidates.

However the smooth screening parametrization employed in Ref. \cite{Vij04},
suggested from old lattice data \cite{Bor89}, does not keep a direct
relation to more recent lattice results \cite{Bal05} pointing out a rather
abrupt string breaking (rehadronization) transition in the potential. Here
we deepen this relation. To this purpose the comparison with a sharply
saturated potential is useful. We analyze the resulting baryon spectrum for
both (smooth and sharp) models and compare the resulting
dominant configurations for any angular momentum. We
generate patterns of quantum numbers and associate to them spectral
regularities observed in the data. From them we are able to make 
purely phenomenological detailed predictions for resonances above 2.3
GeV, many of them not experimentally identified yet. From data and our
predictions we identify $N-\Delta $ degeneracies and $N$ parity series that
have a dynamical understanding within our quark model 
framework.

These contents are organized as follows. In Sec. \ref{secII} we introduce
and discuss the models used and their results for the nonstrange baryon
spectra. The analysis of the dominant contributions and the subsequent
derivation of spectral patterns is carried out in Sec. \ref{secIII}.
Finally, in Sec. \ref{secIV} we summarize our main findings.

\section{The model}
\label{secII}

In the absence of sea quarks lattice QCD predicts the static heavy
quark-heavy antiquark ($Q\overline{Q}$) interacting potential to rise
linearly with the quark-antiquark distance \cite{Bal01}. Unquenched (valence
+ sea quarks) lattice calculations point out the existence of string
breaking in QCD. This is, the static $Q\overline{Q}$ force becomes screened
by intermediate light $q\overline{q}$ pairs, so that when the $Q\overline{Q}$
distance exceeds a critical value (the breaking distance) the $Q\overline{Q}$
potential saturates to a constant (saturation energy). In the 1980's a
smooth screened potential was proposed to parametrize this effect \cite%
{Bor89}. In the last decade such a parametrization has been implemented in
valence quark models in an attempt to examine the consequences of string
breaking in heavy quarkonia as well as in the nonstrange baryon spectra \cite%
{Vij04,Zha93,Din95,Bri00,Gon03}.

The simplest quark-quark screened potential, containing confinement and one
gluon exchange terms, reads (the factor $1/2$ has been explicitly written
since it reflects the difference between the quark-quark and quark-antiquark
interactions):

\begin{equation}
V(r_{ij})=\frac{1}{2}\left[ \overline{\sigma }r_{ij}-\frac{\overline{\kappa }%
}{r_{ij}}+\frac{\hbar ^{2}\overline{\kappa _{\sigma }}}{m_{i}m_{j}c^{2}}%
\frac{e^{-r_{ij}/\overline{r_{0}}}}{\overline{r_{0}}^{2}r_{ij}}(\vec{\sigma
_{i}}\cdot \vec{\sigma _{j}})\right] \left( \frac{1-e^{-\mu \,\,r_{ij}}}{\mu
\,\,r_{ij}}\right) +\frac{\overline{M_{0}}}{3}  \label{eq1}
\end{equation}
where $r_{ij}$ is the interquark distance, $m_{i,j}$ the masses of the
constituent quarks, $\vec{\sigma}_{i,j}$ the spin Pauli operators, and $%
\overline{M_{0}}$ is a constant to fix the ground state nucleon mass. The
screening multiplicative factor appears between parenthesis on the right
hand side. $\mu $, the screening parameter, is the inverse of the saturation
distance and its effective value is fitted together with the other
parameters, $\overline{\sigma }$, $\overline{\kappa }$, and $\overline{%
\kappa _{\sigma }}$, to the spectrum ($\mu^{-1} \simeq 1.4$ fm for heavy
quarkonia and $\mu^{-1} \simeq 1$ fm for nonstrange baryons) \cite%
{Vij04,Gon03}.

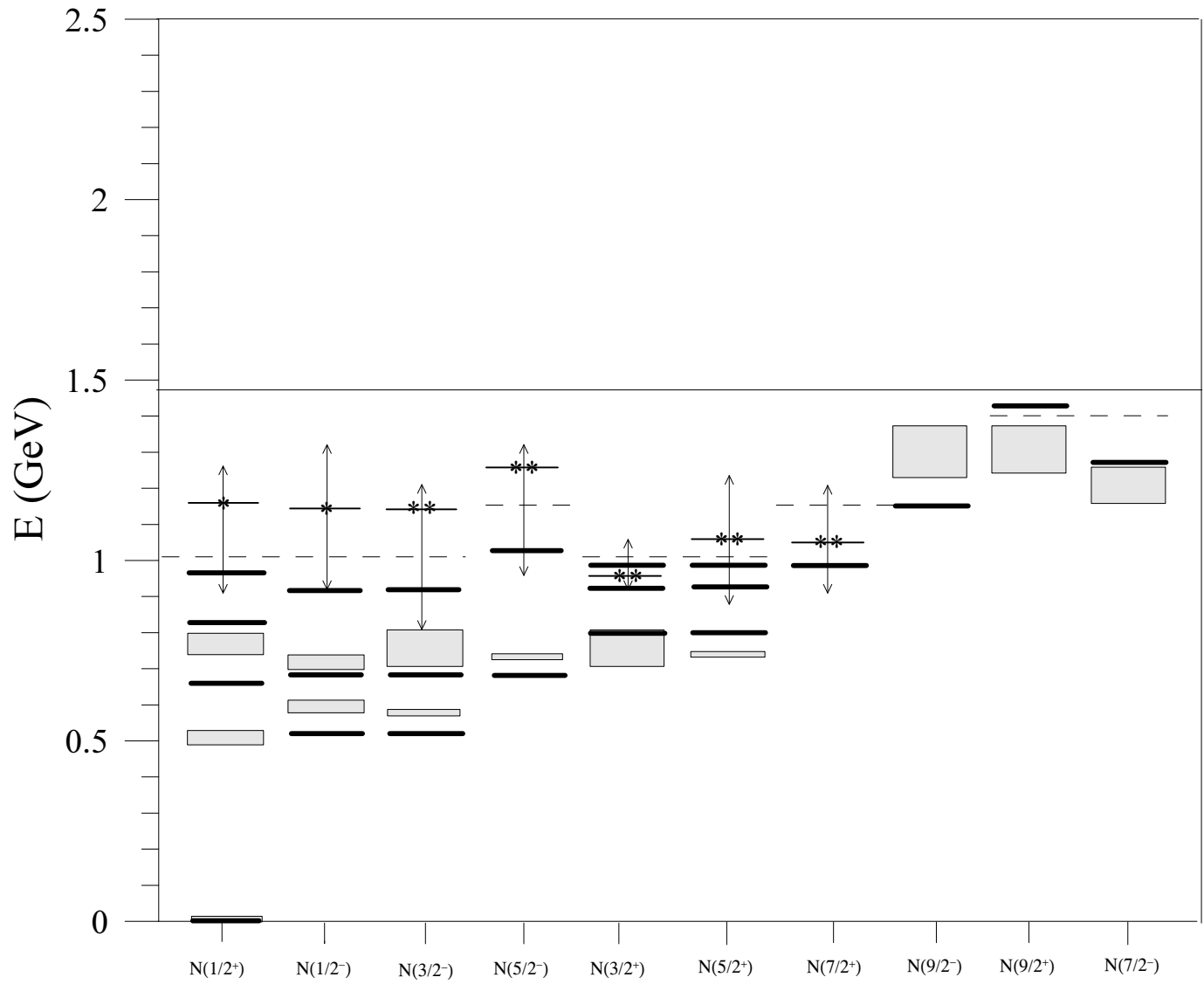
\begin{figure}[tbp]
\caption{ Relative energy nucleon spectrum for the screened potential of Eq.
(\protect\ref{eq1}) with the parameters of Ref. \protect\cite{Vij04}. The
thick solid lines represent our results. The shaded region, whose size
stands for the experimental uncertainty, represents the experimental data
for those states cataloged as $(\ast \ast \ast )$ or $(\ast \ast \ast \ast )$
states in the Particle Data Book \protect\cite{Eid04}. Experimental data
cataloged as $(\ast )$ or $(\ast \ast )$ states are shown by short thin
solid lines with stars over them and by vertical lines with arrows standing
for the experimental uncertainties. Finally, we show by a dashed line the $%
1q $ ionization threshold and by a long thin solid line the total threshold.}
\label{f1}
\end{figure}

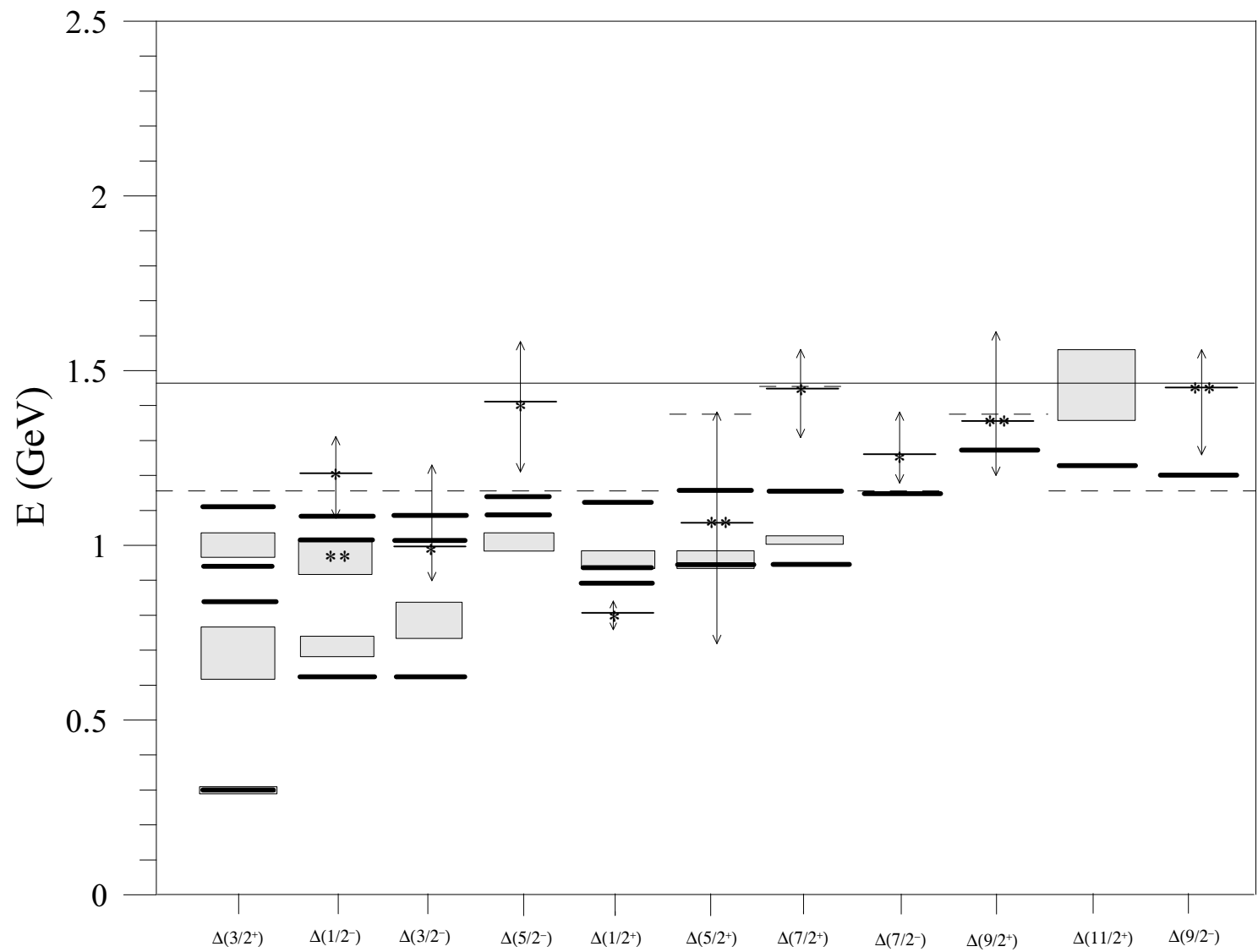
\begin{figure}[tbp]
\caption{Same as Fig. \protect\ref{f1} for $\Delta$ states.}
\label{f2}
\end{figure}

In the heavy meson sector a good description of bottomonium (the only truly
nonrelativistic quarkonia) is obtained. For nonstrange baryons the model
predicts quite approximately the number and ordering of the experimental
states up to a mass of $2.3$ GeV. For the sake of completeness we show
improved results (with more channels entering in the description) from Ref. 
\cite{Vij04} in Figs. \ref{f1} and \ref{f2}. Let us note that the ordering
discrepancies, related to the relative position of the first excited states
of positive and negative parity in $N$ and $\Delta $, are endemic in this
kind of quark model treatments; the number discrepancy related to the
presence of two, instead of one, excited states for $N(3/2^{+})$ and $%
N(5/2^{+})$ might have to do with the presence of two degenerate excitations
at the experimental level of precision.

More recent lattice QCD calculations \cite{Bal05} show that the $Q\overline{Q%
}$ potential saturates sharply for a breaking distance of the order of 1.25
fm corresponding to a saturation energy of about twice the $B$ meson ($Q%
\overline{q})$ mass, indicating that the formation of two heavy-light
subsystems ($B,\overline{B})$ is energetically favored. This information has
been implemented in a quark model scheme \cite{Swa05} showing that, as a
consequence of coupled channels above the physical thresholds (corresponding
to the opening of decay channels), the description becomes progressively
less accurate high in the spectrum. Moreover the mixing with the continuum
can also modify the short-range part of the interaction. Nonetheless an
effective (renormalized) nonscreened potential continues being useful up to
energies not too far above the lowest physical threshold.

These results provide some understanding of the successful application of
the smooth screened potential of Eq. (\ref{eq1}) in Refs. \cite{Vij04} and 
\cite{Gon03}. For bottomonium, the maximum value of the potential, $%
\overline{\sigma }/\mu =2070$ MeV (the $b$ mass has been chosen so that the
constant term in the potential vanishes), translates into an upper limit for
the energy of bound states of $2m_{b}+\overline{\sigma }/\mu =11400$ MeV 
\cite{Gon03}, far above the lowest physical threshold, assuring that the
potential does not differ much from a nonscreened one (corresponding to Eq. (%
\ref{eq1}) without the multiplicative screening factor) for the
experimentally known states. As a matter of fact one can alternatively use
for bottomonium a (linear + coulomb) potential sharply saturated at an
energy just above the highest resonance experimentally known,

\begin{equation}
V(r_{ij})=\left\{ \matrix{ \sigma r_{ij}-\kappa /r_{ij} & & r_{ij} < r_{sat}
\cr \sigma r_{sat}-\kappa /r_{sat} & & r_{ij} \geq r_{sat}}\right. \,,
\end{equation}%
and maintain the quality of the description \cite{Gon05}. Let us note that
this is the closest approach one can do to a nonscreened potential if one
wants to extend its use high in energy since the sharp saturation prevents a
proliferation of predicted states without experimental counterpart. We
should then keep in mind that the fitted saturation distance, $r_{sat}$,
should not be identified with the lattice breaking distance at the physical
threshold but as an effective parameter to establish the limit of
applicability of the model (in fact its fitted value, 1.76 fm, differs a
30\% of the breaking value 1.25 fm).

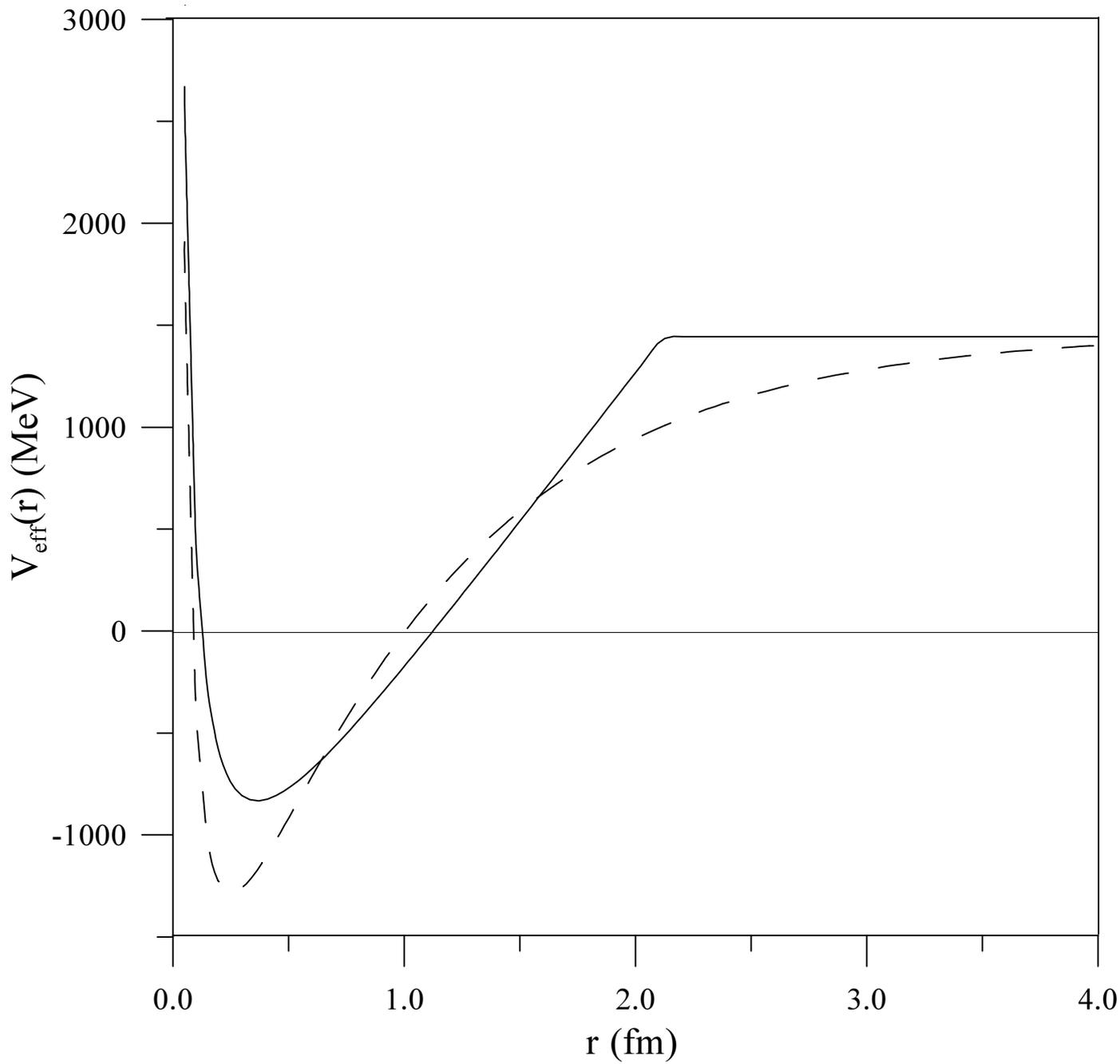
\begin{figure}[tbp]
\caption{Effective interaction, $3\, V(r_{ij}) - E_N$, for the potentials $%
V(r_{ij})$ of Eq. (\protect\ref{eq1}), dashed line, and Eq. (\protect\ref%
{eq2}), solid line, for a two-particle spin 1 state. $E_N$ stands for the
eigenvalue of the nucleon ground state with the corresponding potential.}
\label{f3}
\end{figure}

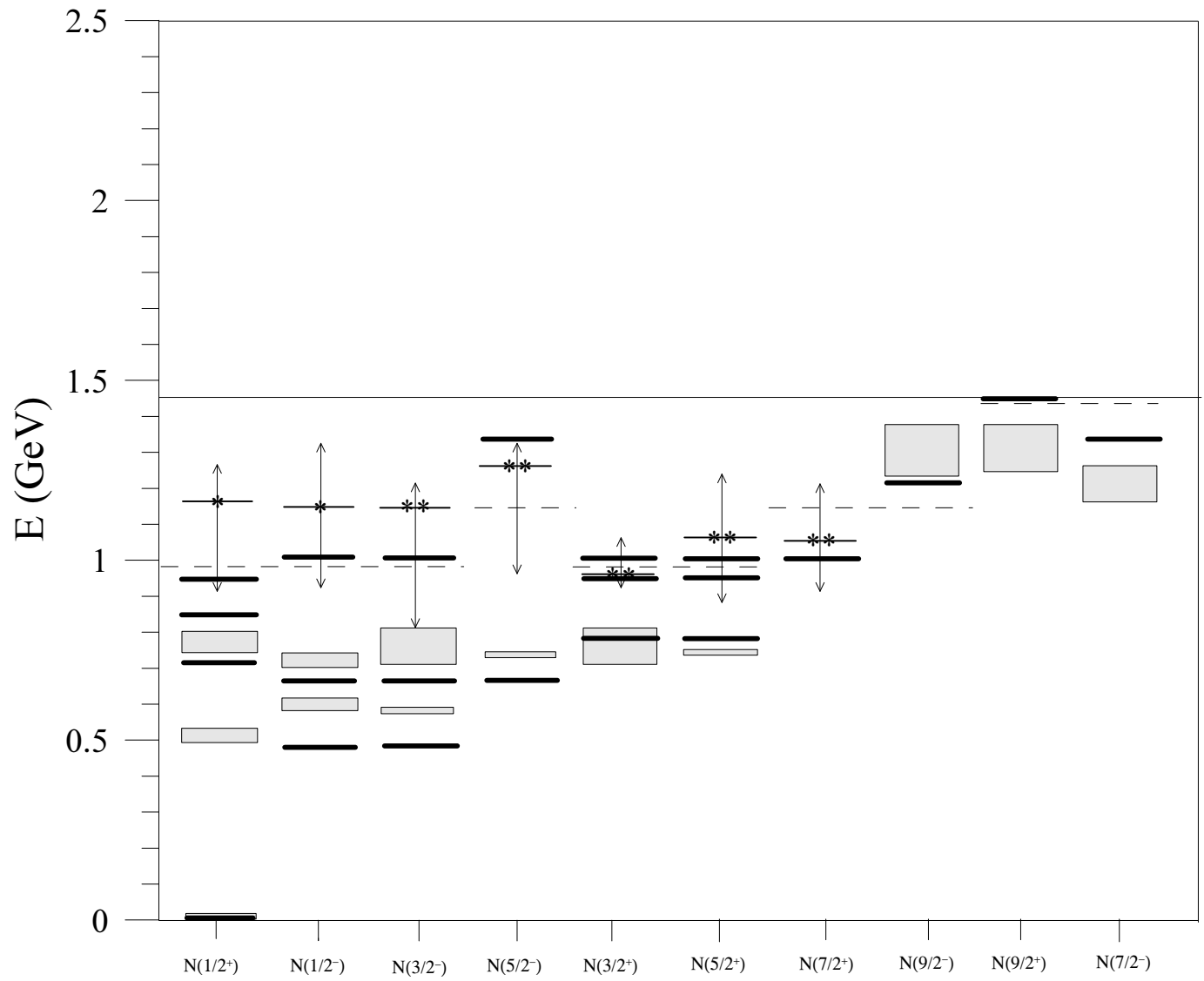
\begin{figure}[tbp]
\caption{Same as Fig. \protect\ref{f1} for the screened potential of Eq. (%
\protect\ref{eq2}) with the parameters of Table \protect\ref{t1}.}
\label{f4}
\end{figure}

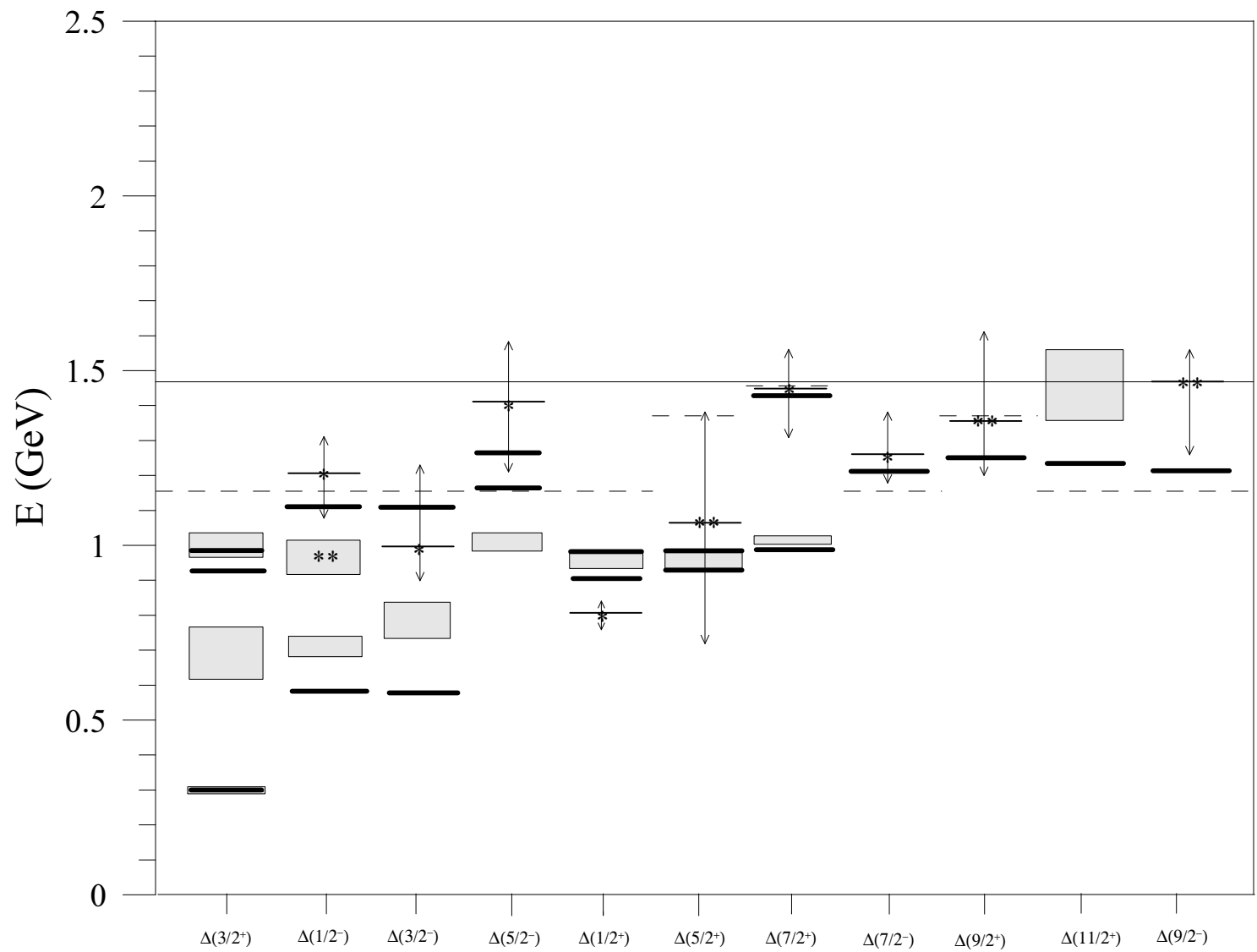
\begin{figure}[tbp]
\caption{ Same as Fig. \protect\ref{f4} for $\Delta$ states.}
\label{f5}
\end{figure}

In the nonstrange baryon case, the maximum possible value of the energy of
bound states for the screened potential in Ref. \cite{Vij04}, $3m_{q}+%
\overline{M_{0}}+3\overline{\sigma }/(2\mu )\simeq 2.4$ GeV , just above the
highest experimental resonance catalogued as four stars by the Particle Data
Group (PDG) \cite{Eid04}, $\Delta (11/2^{+})$ at 2420 MeV, reflects the
difficulty of pushing the applicability limit of the model beyond this
energy. It is interesting to make again a comparison with a sharply
saturated potential of the form:

\begin{equation}
V(r_{ij})=\left\{\matrix{V_{sr}(r_{ij}) & & r_{ij} < r_{sat} \cr 
V_{sr}(r_{sat}) & & r_{ij} \geq r_{sat}}\right. \, ,  \label{eq2}
\end{equation}
where 
\begin{equation}
V_{sr}(r_{ij})=\frac{1}{2}\left[ \sigma r_{ij}- \frac{\kappa }{r_{ij}}+\frac{%
\hbar ^{2}\kappa _{\sigma }} { m_{i}m_{j}c^{2}}\frac{e^{-r_{ij}/r_{0}}}{%
r_{0} ^{2}r_{ij}}(\vec{\sigma _{i}}\cdot \vec{\sigma _{j}})\right] + \frac{%
M_0}{3}
\end{equation}

\noindent whose parameters, given in Table \ref{t1}, are fitted to get a
global description of the nonstrange baryon spectrum while choosing $r_{sat}$
to keep the same value for the saturation energy $3m_{q}+M_{0}+3\sigma
r_{sat}/2\simeq 2.4$ GeV (splitting energy of the three pairs of quarks). We
draw this sharp potential and the smooth one in Fig. \ref{f3}. The
calculation of the spectrum proceeds exactly in the same manner as in Ref. 
\cite{Vij04}, to which we refer for technical details. Let us only remind
for the moment that two different calculational methods: hyperspherical
harmonic expansion (HH) and Faddeev, are employed and that convergence of
the results is required. The resulting spectrum is shown in Figs. \ref{f4}
and \ref{f5}.

It is worth to remark that the presence, in the three-body problem, of
two-body thresholds (for only one quark to be released, values quoted in
Table \ref{t2}), apart from the absolute three-body one (saturation energy),
may represent further constraints in the applicability limit of the model to
any particular channel. As in Ref. \cite{Vij04} we have also included the
predicted states close above the thresholds.

The comparison of Figs. \ref{f4} and \ref{f5} with Figs. \ref{f1} and \ref%
{f2} shows that the sharp potential tends quite generally to push upward the
highest energy states. However, as we can check there is little difference
concerning the number and ordering of states. Both models give them quite
reasonably. In other words the screened potential is quite similar,
concerning these results, to the closest physical approach to a nonscreened 
potential, represented by the sharp interaction, that takes
effectively into account the effect of the baryon decay to open channels in
order to select the observed resonances.

Let us notice further that the sharp potential model permits to establish in
a direct way the subtle connection between the lowest two-body (2bt) and
three-body thresholds (3bt): if the 3bt is increased (changing for example $%
r_{sat})$ and one refits the spin-spin and coulomb terms to recover the
experimental $\Delta -N$ mass difference, the 2bt is increased, almost
linearly, in correspondence. Otherwise said, the 2bt (3bt) plus the $\Delta
-N$ mass difference fixes the 3bt (2bt). This is a realization of the idea
that the mixing with the continuum (at least partially contained in the
effective thresholds) affects the whole (in particular the short) range of
the interaction. Moreover, as the lowest 2bt corresponds to the lowest
angular momentum state, the experimental presence of a well established high
energy-low angular momentum state can be used as a lower bound to fix the
lowest 2bt. Then this fixing procedure determines the maximum energy at
which a higher angular momentum state can be predicted within the range of
applicability of the model (a fine tune of this process has been carried out
to get the spectrum).

\section{Spectral patterns}
\label{secIII}

As shown in Figs. \ref{f1}, \ref{f2}, \ref{f4} and \ref{f5}, the predicted
nonstrange baryon spectra are quite analogous despite the difference between
the two potentials. Furthermore the dominant configurations entering any $%
J^{P}$ ground state and its first nonradial excitation are exactly the same
for both potentials (this is not the case for higher excitations as
discussed below) giving for them similar quantitative values that differ at
most a 13\% from data (within the experimental uncertainties). Hence it
makes sense to proceed, from our dynamical models, to a
quantum number assignement for these states. From the resulting
patterns and from the experimental values of the masses of the known
resonances we shall make predictions for higher energy states and we shall
justify (as implied by our dynamical model) approximate degeneracies
and the partial appearance of approximate parity doublets.

\subsection{$J^{P}$ ground states}
\label{III.A}

To express the spatial part of the dominant $J^{P}$ ground state
configurations we shall use the hyperspherical harmonic notation, i.e., the
quantum numbers $(K,L(\ell ,\lambda ),Symmetry).$ The so-called great
orbital,\ $K,$ defines the parity of the state, $P=(-)^{K},$ and its
centrifugal barrier energy, $\frac{{\cal L}({\cal L}+1)}{2m\left\langle \rho
^{2}\right\rangle }$ (${\cal L}=K+\frac{3}{2}$, $\rho $: hyperrradius)$.$ $%
\ell $ is the orbital angular momentum of a pair of quarks, $\lambda $
stands for the orbital angular momentum of the third quark with respect to
the center of mass of this pair and $L$ ($\vec{L}=\vec{\ell}+\vec{\lambda})$
is the total orbital angular momentum. Alternatively one can write the
parity as $P=(-)^{\ell +\lambda }$, since $K-\ell -\lambda ={\rm even}\geq 0$
is always satisfied. $Symmetry$ specifies the spatial symmetry $(\left[ 3%
\right] :$ symmetric, $\left[ 21\right] :$ mixed, $\left[ 111\right] :$
antisymmetric$)$ which combines to the spin, $S,$ and isospin, $T,$
symmetries $(S,T=3/2:$ symmetric; $S,T=1/2:$ mixed) to have a symmetric wave
function (the color part is antisymmetric). More precisely, $T=1/2$ for $N$
and $T=3/2$ for $\Delta $, hence the spatial-spin wave function must be
mixed for $N$ and symmetric for $\Delta $.

It is convenient to group the ground states corresponding to the same
dominant configuration in our model as done in Tables \ref{t3} and \ref{t4}
(note that above 2.4 GeV the dominant configurations correspond to an
extrapolation of the pattern).

As a first general rule we have: 
\begin{equation}
P=(-)^{L}\, ,
\end{equation}
for the dominant configuration in $J^{P}$ ground states. This comes, except
for $\Delta (1/2^{+})$, from $(\ell=0,\lambda )$ and/or $(\ell ,\lambda =0)$
components for which $L=\ell +\lambda$. For $\Delta (1/2^{+})$ one has $L=0$
with $\ell=1=\lambda$ instead of $\ell=0=\lambda $ due to the more
attractive spin-spin interaction.

The second general rule, of general validity for states with $J\geq 5/2$,
can be formulated as follows:

\begin{equation}
J\geq 5/2\text{ }:\text{ \ }J=L+S\,\,\,with\,\,\,L\,\,\,\,
minimal\,.  \label{ex3}
\end{equation}%
For states with lower total angular momentum such rule is not always
verified. In particular, it is not fulfilled by the $N(3/2^{+})$ and $%
N,\Delta (1/2^{-})$. For $N(3/2^{+})$ the rule would prescribe $(L=0,S=3/2)$%
, hence spin symmetric and consequently spatially mixed. Dynamically a
spatially symmetric $(L=2,S=1/2)$ configuration, with a spin-spin attractive
contribution, is favoured. For $N,\Delta (1/2^{-})$, according to the parity
expression, $L=$ odd and Eq. (\ref{ex3}) cannot be satisfied. In all these
exceptional cases $J=L-S,$ with $S$ minimal, provides the minimum energy.
Otherwise said, at the level of dominant configurations, the $N(3/2^{+})$
and the $N(5/2^{+})$ are degenerate as well as the $N,\Delta (1/2^{-})$ and
the $N,\Delta (3/2^{-})$, respectively. These degeneracies are quite well
satisfied by data, see Tables \ref{t3} and \ref{t4}.

These two rules, complemented with the prescription of having the minimum
value of $K$ (i.e., the minimum centrifugal barrier) satisfying the symmetry
requirements, express the minimum energy $J^{P}$ configuration coming out
from our dynamical model.

It is worth to mention that the same spectral patterns and rules
could be also derived from a careful look at data in order to apply a
general quark model multiplet structure as done in Ref. \cite{Kle02}.
However, in this case the lack of an underlying
dynamics prevents, as pointed by the author, a justification for the
phenomenological rules derived.

\subsection{Positive parity states}

For positive parity states the emerging picture consists of two sets
(spatially symmetric and mixed corresponding to the upper and lower parts of
Table \ref{t3}, respectively) of $N^{\prime }s$ and $\Delta ^{\prime }s$
grouped according to the same dominant configurations. Each $\Delta $ state,
except for the $\Delta (1/2^{+})$, has a nucleon correspondence. For $\Delta
(1/2^{+})$ its {\it natural} nucleon partner should be a $N(3/2^{+})$ with $%
(L=0,S=3/2)$ which, as mentioned above, is dynamically unfavored. Actually,
our model predicts for $N(3/2^{+})$ $(L=2,S=1/2)$ a lower mass than for $%
\Delta (1/2^{+})$ as it is the experimental case.

It is important to realize that the percentage of the dominant configuration
in the sets varies: while for the upper part it can be quite different for $%
N $ and $\Delta $ partners (for example $62\%$ of $(2,2,\left[ 3\right] )$
for $N(5/2^{+})$ and 99\% for $\Delta (7/2^{+})),$ for the lower part it is
quite approximately the same and very close to $100\%$ (for instance 99\% of 
$(2,2,\left[ 21\right] )$ for $\Delta (5/2^{+})$ and $99\%$ for $%
N(7/2^{+})). $ This has to do with the fact that for the lower part series
only a spatially mixed wave function is allowed whereas for the upper part $%
N $ series there are, except for $N(1/2^{+})$, two possibilities open for
the same $(K,L)$ values: spatially symmetric (spin-isospin symmetric) and
spatially mixed (spin-isospin mixed).

For the upper part set the lowest lying states, $N(1/2^{+})$ and $\Delta
(3/2^{+})$, with a dominant configuration $(0,0,\left[ 3\right] )$, have an
energy difference of 292 MeV, entirely due in our model to the spin-spin
potential, attractive for $S=1/2$ and repulsive for $S=3/2,$ since for both
states the percentage of the dominant configuration is almost the same ($98\%
$ for $N(1/2^{+})$ and $100\%$ for $\Delta (3/2^{+}).$ Let us keep in mind
that {\it chiral} contributions from meson exchanges are taken into account
through the effective quark-gluon coupling constant fitted to the $\Delta
(3/2^{+})-N(1/2^{+})$ mass difference. For the next dominant configuration,
containing the $N(5/2^{+})$ and the $\Delta (7/2^{+}),$ whose masses are
quite well reproduced by our model, one has to consider the configuration
mixing effect in $N(5/2^{+})$ ($62\%$ of $(2,2,\left[ 3\right] )$ and $34\%$
of $(2,2,\left[ 21\right] ))$. Let us realize though that both
configurations would give the same energy for an hypercentral (only $\rho -$%
dependent) potential. As our (confinement + coulomb) potential is quite
close to an hypercentral one, the difference between $N$ and $\Delta $
partners is again due in our model to the spin-spin interaction. We predict
199 MeV. This means a slow variation of the spin-spin contribution when
increasing $K$ and $L$ for spatially symmetric states (actually the $62\%$
of the $\Delta (3/2^{+})-N(1/2^{+})$ mass difference is $181$ MeV). The
experimental $\Delta -N$ energy difference is bigger, $\Delta
(7/2^{+})-N(5/2^{+})\simeq $ 270 MeV, but also seem to vary slowly: $\Delta
(11/2^{+})-N(9/2^{+})\simeq 200$ MeV, $\Delta (15/2^{+})-N(13/2^{+})\simeq
250$ MeV. This may be reflecting a larger probability for the spatially
symmetric component than given by our model (let us point in addition that
the more cumbersome numerical procedure for $J\geq 9/2$ states due to the
presence of thresholds, makes less accurate our predictions, indeed the
calculated energies for $N(9/2^{+})$ and $\Delta (11/2^{+})$ are inverted
respect to data). These quantitative differences should be expected
given the shortcomings of our dynamical model. However the spectral pattern
derived is very useful to make manifest a regularity in the data:
the experimental $N$ and $\Delta $ energy steps when going from one
dominant configuration to the next one remain also constant about 500 MeV:
$N(9/2^{+})-N(5/2^{+})\simeq 540$ MeV, $\Delta (11/2^{+})-\Delta
(7/2^{+})\simeq 470$ MeV, $N(13/2^{+})-N(9/2^{+})\simeq 480$ MeV, $\Delta
(15/2^{+})-\Delta (11/2^{+})\simeq 530$ MeV (this energy step can be
associated, from our model, to $\Delta K=\Delta L=2$). The extension of
this pattern to further steps allows us to predict that a $N(17/2^{+})$, if
existing identifiable experimental resonances at such high energies, would
have a mass of 3200 MeV.

Regarding the lower part set we should notice that the same configuration
percentage for $N^{\prime }s$ and $\Delta ^{\prime }s$ would imply a
degeneracy, term by term, if only spatial (neither spin nor isospin
dependence) interactions were considered, $N(7/2^{+}):\Delta (5/2^{+})$; $%
N(11/2^{+}):\Delta (9/2^{+})$, ... As a matter of fact, as for this set the
spin-spin interaction (the only non-spatial in our model) gives little
contribution, due to its short-range character and the spatial mixed
symmetry involved (with $K=L\geq 2)$, the difference between the $N(7/2^{+})$
and $\Delta (5/2^{+})$ masses (the only states clearly below threshold) is
in our model less than $5\%$ what seems to be corroborated by data (this
could be an indication that isospin dependent interactions do not contribute
significantly as well). Concerning the energy step we expect, from our
analysis above ($\Delta K=\Delta L=2)$ and the scarce data ($\Delta
(9/2^{+})-\Delta (5/2^{+})\simeq 400$ MeV) an approximate constant value
around 400$-$500 MeV. From it there would appear a $N(11/2^{+})$ at about
2450 MeV (the energy of its two stars $\Delta (9/2^{+})$ partner is
experimentally in the interval 2200$-$2500 MeV), a $\Delta (13/2^{+})$ and $%
N(15/2^{+})$ at about 2900 MeV and a $\Delta (17/2^{+})$ around 3300 MeV.

All predictions of our model up to 3 GeV, from this subsection and next
subsections \ref{subC} and \ref{subE}, are summed up in Table \ref{t7}. 

\subsection{Negative parity states}
\label{subC}

For negative parity states the results are organized in Table \ref{t4}. The
upper part corresponds to spatially mixed configurations for which we
expect, from our model, a rapid decrement in the spin-spin energy
difference between $N$ and $\Delta $ partners when going from $K=L=1$ to $%
K=L\geq 3$ and, as a consequence, all the states with the same dominant
configuration should become close in mass. The data seem to follow this
prescription: mass differences of 180 MeV at most for $K=L=1$, 60 MeV for $%
K=L=3...$. Again the spectral pattern makes manifest the energy step
(400$-$500 MeV) regularity in the data: ($N(9/2^{-})-N(5/2^{-})=575$ MeV; $%
\Delta (7/2^{-})-\Delta (3/2^{-})=500$ MeV, $N(11/2^{-})-N(7/2^{-})=410$
MeV) as it corresponds to $\Delta K=\Delta L=2$. Therefore one would expect
a $N(13/2^{-})$ and a $\Delta (11/2^{-})$ at about 2650 MeV and a 
$N(15/2^{-})$, a $\Delta (15/2^{-})$ and a $N(17/2^{-})$ at about 3050 MeV.

The lower part of Table \ref{t4} contains an independent series of $\Delta
^{\prime }s$ without $N^{\prime }s$ partners. All states are above their
lowest threshold. From the analysis above we can expect again a step
energy, corresponding to $\Delta K=\Delta L=2,$ of around 400$-$500
MeV that turns out to be compatible with data within the experimental
errors. From it a $\Delta (17/2^{-})$ would be located at approximately 3200
MeV.

\subsection{N-$\Delta$ approximate degeneracies}

According to our model, the masses of the $N(J^{-})$ and $\Delta
(J^{-})$ in the left hand side of Table \ref{t4} should become degenerate as
the effective spin-spin interaction becomes weaker and the potential
approaches an hypercentral one. This seems also to be the experimental
situation.

On the other hand, the $N(J^{+})$ and $\Delta (J^{+})$ on the right
hand side of Table \ref{t3} should become degenerate as the spin-spin
contribution for $S=3/2$ (significantly smaller than for $S=1/2$ in
spatially symmetric states) is not quantitatively much different for
spatially symmetric and mixed states. This is confirmed by the scarce
data available ($J^P=7/2^+$).

These results, from data and our predictions, are picked up in Table \ref{t5}.
We can express its content in a schematic simplified manner through: 
\begin{equation}
N(J^{+,-})\simeq \Delta (J^{+,-})\,\,\,\,\,\,\,for\,\,\,\,\,\,\,J=%
\frac{4n+3}{2}\,\,\,\,\,\,\,with\,\,\,\,\,\,\,n=1,2,3...
\label{eq7}
\end{equation}

\noindent
Let us realize that for positive parity Eq. (\ref{eq7}) corresponds to a rule
derived in Ref. \cite{Kle02} from a phenomenological mass formula
obtained from the assumption of linear Regge trajectories. 

\subsection{Excited states}
\label{subE}

For excited states in general the situation is much more cumbersome. Though
both models, smooth and sharp, provide a similar pattern a more careful
analysis of the configurations entering the states makes clear significant
differences. For instance the fourth $N(1/2^{+})$ state has for the smooth
potential a $(2,2,\left[ 21\right] ),S=3/2$ component degenerate (at the
level of precision of the model) with the $(2,1,[111]),$ $S=1/2$, while the
same state corresponds uniquely to the last configuration for the sharp
potential. For $\Delta (3/2^{+})$ in the sharp case, the radial excitation
of $(0,0,\left[ 3\right] ),S=3/2$ and the $(2,2,\left[ 21\right] ),S=1/2$
configurations are degenerate, whereas in the smooth case, the radial
excitation of $(0,0,\left[ 3\right] ),S=3/2$ is the first excited state and
it is not degenerate, etc. Moreover, the first radial excitations of $\Delta
(3/2^{+})$ at 1600 MeV and $N(1/2^{+})$ at 1440 MeV are badly predicted. In
contrast to this inadequacy to deal with the radial excitations and the
strong model dependence of the configurations for higher (second and up)
excitations, our dynamics provides a very simple picture for the first
non-radial excitations that follows the trend of data. The rule, 
coming from the absence of spin-orbit and tensor forces in our model, and
again of general validity for $J\geq 5/2$, is:

\bigskip

{\it The first non-radial excitation of $N,\Delta(J)$ and the ground
state of $N,\Delta(J+1)$ respectively, are almost degenerate.} \hfill (8)

\bigskip

\noindent This rule is satisfied by existing data at the level of $3\%$,
Table \ref{t6}. From now on we will denote by a black dot the first
non-radial excitation of any state. The only significant deviation from the
previous rule, $\Delta (5/2^{-})^{\bullet }(2350):\Delta (7/2^{-})(2200)$ ($%
7\%$ of difference), corresponds to two resonances catalogued as one star by
the PDG. Indeed the degeneracy can be accomplished within the experimental
uncertainties, see Fig. \ref{f5}.

Assuming this approximate degeneracy we can guess, from data and from our 
phenomenological ground state predictions above, the masses for the
first non-radial excited states with a question mark in Table \ref{t6}.
These predictions, up to $\simeq 3$ GeV, are summed up in Table \ref{t7}.

For $J < 5/2$ the rule is only satisfied, according to our model, by $%
N(3/2^{-})$ and $\Delta (3/2^{+})$. For some of the remaining states our
model gives also definite relations between their first non-radial
excitations and other ground or excited states. These relations are quite
reasonably satisfied by the data too: $N(1/2^{-})^{\bullet
}(1650):N(3/2^{-})^{\bullet }(1700):N(5/2^{-})(1675)$, $\Delta
(1/2^{-})^{\bullet }(1900):\Delta (3/2^{-})^{\bullet }(1940)$, $\Delta
(1/2^{+})^{\bullet }(1910):\Delta (5/2^{+})^{\bullet }(2000):\Delta
(7/2^{+})(1950)$.

It is worth to remark that some of our findings in Table \ref{t7} might be
in fact already accomplished by current data although being masked by the
experimental uncertainties (see Figs. \ref{f4} and \ref{f5}). For example
the $N(7/2^{+,-})^{\bullet }$ could be, within theoretical plus experimental
errors, in the $N(7/2^{+,-})$, the $\Delta (7/2^{-})^{\bullet }$ in the $%
\Delta (7/2^{-})$ and so on. We should also realize that some of the 
predicted resonances could have small couplings to formation channels what
would make difficult its detection (see for example Ref. \cite{Cap93}
regarding $N(11/2^+)$, $N(13/2^-)$ and $N(15/2^+)$).

\subsection{Parity series}

A look at Tables \ref{t3}, \ref{t4} and \ref{t7} shows some approximate $%
N(J^{P})$ parity doublets. Thus, for $J\geq 5/2$ the $N^{+}$ ground state
series in the upper part of Table \ref{t3} and the $N^{-}$ ground state
series in the upper part of Table \ref{t4} appear almost degenerate, term by
term, for example $N(5/2^{+})(1680):N(5/2^{-})(1675)$, as shown on the left
hand side of Table \ref{t8}. From our model we can propose a
qualitative systematic for this doubling: the bigger repulsion in the
positive parity state with respect the negative parity one, due to $\Delta
K=1,\Delta L=1$, is compensated by the bigger attraction in a spatially
symmetric $S=1/2$ configuration (i.e., $\Delta S=-1)$. We can quantify these
effects from our data analysis above. We can estimate an increment of the
repulsion energy of $\simeq 250$ MeV (from $\simeq 500$ MeV for $\Delta
K=2,\Delta L=2,$ as previously discussed) and an increment in the
attraction, when going from $S=3/2$ to $S=1/2$ of approximately the same
amount. So we can predict further parity doublings, $N(13/2^{+}):N(13/2^{-})$%
, ... Moreover the application of our preceding rules for the first
non-radial excitations gives rise to an equivalent approximate parity
doublet series, right hand side of Table \ref{t8}.

For $\Delta ^{\prime }s$ ($J\geq 5/2)$ although an inspection to
data in Tables \ref{t3} and \ref{t4} might also suggest some approximate
degeneracies: the $\Delta (J^{+,-})$ ground state series in the lower parts
of Table \ref{t3} and Table \ref{t4} could be almost degenerate, term by
term: $\Delta (5/2^{+})(1905):\Delta (5/2^{-})(1930),$ $\Delta
(9/2^{+})(2300):\Delta (9/2^{-})(2400)...$., we expect our model not
to support this suggestion. Indeed going from the positive to the
corresponding negative parity state we expect the centrifugal repulsion
getting increased $(\Delta K=1)$ and the spin-spin interaction changing from
attractive $(S=1/2)$ to repulsive $(S=3/2)$, hence the $J^{-}$ states should
be higher in mass than the $J^{+}$ ones.

We should realize that the approximate $N$ parity doublets appear
discretely, every 500 MeV. This does not preclude the existence for other
energies of additional ($N$ and/or $\Delta $) parity doublets involving
higher excitations but we cannot carry out for them a trustable analysis.

\section{Summary}
\label{secIV}

We have performed a phenomenological analysis of the nonstrange
baryon spectrum between 2.3 and 3 GeV. Our starting point is a quark model
potential, containing confinement and OGE interactions saturating at 2.4
GeV, that gives the correct number and ordering of known resonances up to
2.3 GeV. The study of the dominant configurations entering the ground state
and the first non-radial excitations provides us with quantum number
spectral patterns. The extrapolation of these patterns to
higher energies allows the prediction, from existing data, of
ground ($J\geq 11/2)$ and first non-radial excited ($J\geq 7/2)$ state
masses in the quite experimentally uncertain energy region from 2.3 to 3
GeV. It is precisely this uncertainty, reflecting the transition to the
continuum, what makes our picture plausible despite the blown up number of 
{\it new} states.

The qualitative adequacy of our model description to data suggests that the
effective (confinement+coulomb+spin-spin) dynamics employed (we have
estimated numerically small contributions of conventional meson
exchange potentials to the mass of $J\geq 5/2$ baryons) contains the
essential ingredients to give account of the observed regularities in the
spectrum. To this respect we should emphasize i) the usefulness of
screening as an effective mechanism in quark potential models to provide an
unambiguous assignment of quantum numbers to states (this is related to the
connection of screening to decay channel effects as shown by lattice
calculations) and ii) the importance of using the hyperspherical harmonic
basis as the more convenient to deal, from the point of view of the physical
interpretation, with the quasi-hypercentral potential considered.

From data and our predictions we find on the one hand approximate $N-\Delta $
degeneracies and on the other hand series of $N$ parity doublets involving
ground or first non-radial excitations exclusively. We get from our 
minimal dynamics some understanding about how these degeneracies come out.

We should finally remark that our results do not rely on any symmetry
assumption in QCD, but on a particular dynamics that provides a spectral
pattern in agreement to data. Since existing data beyond 1.9 GeV are plagued
with large experimental errors we cannot pretend our quantitative
predictions to be very precise. Nonetheless we think our findings can be of
help for the always needed experimental spectral searches in this region as
well as for the theoretical progress in the dynamical interpretation of data.

\acknowledgments
This work has been partially funded by Ministerio de Ciencia y Tecnolog\'{\i}a
under Contract No. FPA2004-05616, by Junta de Castilla y Le\'{o}n under
Contract No. SA-104/04, and by Generalitat Valenciana under Contract No.
GV05/276.

\narrowtext 
\begin{table}[tbp]
\caption{Quark model parameters.}
\label{t1}%
\begin{tabular}{|cccc|}
& $m_u=m_d \,\, ({\rm {MeV}) }$ & 337 &  \\ 
& $r_0 \,\, ({\rm {fm}) }$ & 0.495 &  \\ 
& $r_{sat} \,\, ({\rm {fm}) }$ & 2.12 &  \\ 
& $\kappa \,\, ({\rm {MeV \, fm}) }$ & 10.0 &  \\ 
& $\kappa_\sigma \,\, ({\rm {MeV \, fm}) }$ & 120.0 &  \\ 
& $\sigma \,\, ({\rm {MeV \, fm}^{-1}})$ & 976.56 &  \\ 
& $M_0 \,\, ({\rm {MeV}) }$ & $-$ 1726.78 & 
\end{tabular}%
\end{table}

\begin{table}[tbp]
\caption{One-quark ionization thresholds. We give the energy above the
nucleon ground state mass.
$\ell $ is the orbital angular momentum of a pair of quarks, $s$ its relative
spin and $t$ its isospin.}
\label{t2}%
\begin{tabular}{|cccc|}
& $(\ell,s,t)$ & E (MeV) &  \\ \hline
& $(0,0,0)$ & 984.10 &  \\ 
& $(0,1,1)$ & 1143.33 &  \\ 
& $(1,0,1)$ & 1384.15 &  \\ 
& $(1,1,0)$ & 1416.48 &  \\ 
&  &  & 
\end{tabular}%
\end{table}

\begin{table}[tbp]
\caption{Positive parity $N$ and $\Delta$ ground states for different
dominant spatial-spin configurations up to $\simeq 3$ GeV. 
The assignment of dominant configurations above 2.4 GeV corresponds 
to an educated guess. Experimental data
are from PDG \protect\cite{Eid04}. Stars have been omitted for four-star
resonances. States denoted by a question mark correspond to predicted
resonances that do not appear in the PDG (their predicted masses, also
indicated by a question mark, appear in Table \protect\ref{t7}).}
\label{t3}
\begin{center}
\begin{tabular}{|c|c|c|c|c|c|c|}
$(K,L,Symmetry)$ & $S=1/2$ & \multicolumn{2}{c|}{Model (MeV)} & 
\multicolumn{2}{c|}{Exp. (MeV)} & $S=3/2$ \\ \hline
$(0,0,[3])$ & $N(1/2^+)$ & 940 &  & 940 &  &  \\ \cline{2-7}
&  &  & 1232 &  & 1232 & $\Delta(3/2^+)$ \\ \hline
$(2,2,[3])$ & $N(5/2^+),N(3/2^+)$ & 1722 &  & 1680,1720 &  &  \\ \cline{2-7}
&  &  & 1921 &  & 1950 & $\Delta(7/2^+)$ \\ \hline
$(4,4,[3])$ & $N(9/2^+)$ & 2378 &  & 2220 &  &  \\ \cline{2-7}
&  &  & 2175 &  & 2420 & $\Delta(11/2^+)$ \\ \hline
$(6,6,[3])$ & $N(13/2^+)(**)$ &  &  & 2700 &  &  \\ \cline{2-7}
&  &  &  &  & 2950 & $\Delta(15/2^+)(**)$ \\ \hline\hline
$(2,0,[21])$ & $\Delta(1/2^+)$ &  & 1849 &  & 1750 &  \\ \hline
$(2,2,[21])$ &  & 1938 &  & 1990 &  & $N(7/2^+)(**)$ \\ \cline{2-7}
& $\Delta(5/2^+)$ &  & 1871 &  & 1905 &  \\ \hline
$(4,4,[21])$ &  &  &  & $?$ &  & $N(11/2^+)(?)$ \\ \cline{2-7}
& $\Delta(9/2^+)(**)$ &  & 2193 &  & 2300 &  \\ \hline
$(6,6,[21])$ &  &  &  & $?$ &  & $N(15/2^+)(?)$ \\ \cline{2-7}
& $\Delta(13/2^+)(?)$ &  &  &  & $?$ & 
\end{tabular}%
\end{center}
\end{table}

\begin{table}[tbp]
\caption{Negative parity $N$ and $\Delta $ ground states for different
dominant spatial-spin configurations up to $\simeq 3$ GeV. 
The assignment of dominant configurations above 2.4 GeV 
correspond to an educated guess. Experimental data
are from PDG \protect\cite{Eid04}. Stars have been omitted for four-star
resonances. States denoted by a question mark correspond to predicted
resonances that do not appear in the PDG (their predicted masses, also
indicated by a question mark appear in Table \protect\ref{t7}).}
\label{t4}
\begin{center}
\begin{tabular}{|c|c|c|c|c|c|c|}
$(K,L,Symmetry)$ & $S=1/2$ & \multicolumn{2}{c|}{Model (MeV)} & 
\multicolumn{2}{c|}{Exp. (MeV)} & $S=3/2$ \\ \hline
& $N(3/2^-), N(1/2^-)$ & 1410 &  & 1520,1535 &  &  \\ \cline{2-7}
$(1,1,[21])$ &  & 1596 &  & 1675 &  & $N(5/2^-)$ \\ \cline{2-7}
& $\Delta(3/2^-),\Delta(1/2^-)$ &  & 1517 &  & 1700,1620 &  \\ \hline
& $N(7/2^-)$ & 2275 &  & 2190 &  &  \\ \cline{2-7}
$(3,3,[21])$ &  & 2153 &  & 2250 &  & $N(9/2^-)$ \\ \cline{2-7}
& $\Delta(7/2^-)(*)$ &  & 2153 &  & 2200 &  \\ \hline
& $N(11/2^-)(***)$ &  &  & 2600 &  &  \\ \cline{2-7}
$(5,5,[21])$ &  &  &  & $?$ &  & $N(13/2^-)(?)$ \\ \cline{2-7}
& $\Delta(11/2^-)(?)$ &  &  &  & $?$ &  \\ \hline\hline
$(3,1,[3])$ &  &  & 2114 &  & 1930 & $\Delta(5/2^-)(***)$ \\ \hline
$(5,3,[3])$ &  &  & 2153 &  & 2400 & $\Delta(9/2^-)(**)$ \\ \hline
$(7,5,[3])$ &  &  &  &  & 2750 & $\Delta(13/2^-)(**)$%
\end{tabular}%
\end{center}
\end{table}

\begin{table}[tbp]
\caption{$N$ and $\Delta$ approximate degeneracies up to $\simeq$ 3 GeV.
Experimental masses (in MeV) from PDG \protect\cite{Eid04}. Predicted masses
are signaled by a question mark.}
\label{t5}
\begin{center}
\begin{tabular}{|ccccc|}
& $N$ &  & $\Delta$ &  \\ \hline
& $N(7/2^-)(2190)$ & : & $\Delta(7/2^-)(2200)$ &  \\ 
& $N(11/2^-)(2600)$ & : & $\Delta(11/2^-)(2650?)$ &  \\ \hline
& $N(7/2^+)(1990)$ & : & $\Delta(7/2^+)(1950)$ &  \\ 
& $N(11/2^+)(2450?)$ & : & $\Delta(11/2^+)(2420)$ &  \\ 
& $N(15/2^+)(2900?)$ & : & $\Delta(15/2^+)(2950)$ & 
\end{tabular}%
\end{center}
\end{table}

\begin{table}[tbp]
\caption{Ground and first non-radial excitation (denoted by a black dot)
correspondence according to our model up to $\simeq 3$ GeV. Experimental
masses (in MeV) from PDG \protect\cite{Eid04}. Experimental unknown masses
are signaled by a question mark.}
\label{t6}
\begin{center}
\begin{tabular}{|cccc||cccc|}
& \multicolumn{3}{c||}{$N$} & \multicolumn{3}{c}{$\Delta$} &  \\ \hline
&  &  &  & $\Delta(3/2^+)^\bullet$(1920) & : & $\Delta(5/2^+)$(1905) &  \\ 
& $N(5/2^+)^\bullet$(2000) & : & $N(7/2^+)$(1990) & $\Delta(5/2^+)^\bullet$%
(2000) & : & $\Delta(7/2^+)$(1950) &  \\ 
& $N(7/2^+)^\bullet$(?) & : & $N(9/2^+)$(2220) & $\Delta(7/2^+)^\bullet$%
(2390) & : & $\Delta(9/2^+)$(2300) &  \\ 
& $N(9/2^+)^\bullet$(?) & : & $N(11/2^+)$(?) & $\Delta(9/2^+)^\bullet$(?) & :
& $\Delta(11/2^+)$(2420) &  \\ 
& $N(11/2^+)^\bullet$(?) & : & $N(13/2^+)$(2700) & $\Delta(11/2^+)^\bullet$%
(?) & : & $\Delta(13/2^+)$(?) &  \\ 
& $N(13/2^+)^\bullet$(?) & : & $N(15/2^+)$(?) & $\Delta(13/2^+)^\bullet$(?)
& : & $\Delta(15/2^+)$(2950) &  \\ \hline\hline
& $N(3/2^-)^\bullet$(1700) & : & $N(5/2^-)$(1675) &  &  &  &  \\ 
& $N(5/2^-)^\bullet$(2200) & : & $N(7/2^-)$(2190) & $\Delta(5/2^-)^\bullet$%
(2350) & : & $\Delta(7/2^-)$(2200) &  \\ 
& $N(7/2^-)^\bullet$(?) & : & $N(9/2^-)$(2250) & $\Delta(7/2^-)^\bullet$(?)
& : & $\Delta(9/2^-)$(2400) &  \\ 
& $N(9/2^-)^\bullet$(?) & : & $N(11/2^-)$(2600) & $\Delta(9/2^-)^\bullet$(?)
& : & $\Delta(11/2^-)$(?) &  \\ 
& $N(11/2^-)^\bullet$(?) & : & $N(13/2^-)$(?) & $\Delta(11/2^-)^\bullet$(?)
& : & $\Delta(13/2^-)$(2750) & 
\end{tabular}%
\end{center}
\end{table}

\begin{table}[tbp]
\caption{$N$ and $\Delta $ predicted states in the interval $[2.2,3.0]$ MeV.
We denote by a black dot the first non-radial excitation.}
\label{t7}
\begin{center}
\begin{tabular}{|c|c|c||c|c|}
& \multicolumn{2}{c||}{$N$} & \multicolumn{2}{c|}{$\Delta$} \\ \hline
$J=7/2$ & $N(7/2^+)^\bullet$(2220) & $N(7/2^-)^\bullet$(2250) &  & $%
\Delta(7/2^-)^\bullet$(2400) \\ 
$J=9/2$ & $N(9/2^+)^\bullet$(2450) & $N(9/2^-)^\bullet$(2600) & $%
\Delta(9/2^+)^\bullet$(2420) & $\Delta(9/2^-)^\bullet$(2650) \\ 
$J=11/2$ & $N(11/2^+)$(2450) &  &  & $\Delta(11/2^-)$(2650) \\ 
& $N(11/2^+)^\bullet$(2700) & $N(11/2^-)^\bullet$(2650) & $%
\Delta(11/2^+)^\bullet$(2850) & $\Delta(11/2^-)^\bullet$(2750) \\ 
$J=13/2$ &  & $N(13/2^-)$(2650) & $\Delta(13/2^+)$(2850) &  \\ 
& $N(13/2^+)^\bullet$(2900) &  & $\Delta(13/2^+)^\bullet$(2950) &  \\ 
$J=15/2$ & $N(15/2^+)$(2900) &  &  & 
\end{tabular}%
\end{center}
\end{table}

\begin{table}[tbp]
\caption{Dynamical parity doublets from our analysis up to $\simeq 3$ GeV.
Experimental masses (in MeV) from PDG \protect\cite{Eid04}. Predicted masses
are signaled by a question mark. The black dot indicates the first
non-radial excitation.}
\label{t8}
\begin{center}
\begin{tabular}{|ccc||ccc|}
\multicolumn{6}{|c|}{$N$} \\ \hline
\multicolumn{3}{|c||}{\rm Ground states} & \multicolumn{3}{c|}{\rm First
non-radial excitation} \\ \hline
$N(5/2^+)$(1680) & : & $N(5/2^-)$(1675) &  &  &  \\ 
&  &  & $N(7/2^+)^\bullet$(2220?) & : & $N(7/2^-)^\bullet$(2250?) \\ 
$N(9/2^+)$(2220) & : & $N(9/2^-)$(2250) &  &  &  \\ 
&  &  & $N(11/2^+)^\bullet$(2700?) & : & $N(11/2^-)^\bullet$(2650?) \\ 
$N(13/2^+)$(2700) & : & $N(13/2^-)$(2650?) &  &  & 
\end{tabular}%
\end{center}
\end{table}

\end{document}